\documentclass[twocolumn,prd,preprintnumbers,amsmath,amssymb]{revtex4}
\pdfoutput=1
\usepackage{graphicx}
\usepackage{amssymb}
\usepackage{psfrag}
\let\oldhat\hat
\renewcommand{\vec}[1]{\mathbf{#1}}
\renewcommand{\hat}[1]{\oldhat{\mathbf{#1}}}

\begin{document}

\title{Computing General Relativistic effects from Newtonian N-body simulations:\\
Frame dragging in the post-Friedmann approach}

\author{Marco Bruni}\email{marco.bruni@port.ac.uk} \author{Daniel B.~Thomas}\email{daniel.b.thomas@port.ac.uk} \author{David Wands}\email{david.wands@port.ac.uk}
\affiliation{Institute of Cosmology and Gravitation, University of Portsmouth, UK}
\date{5 June 2013}

\begin{abstract}
\hspace{-0.33cm}We present the first calculation of an intrinsically relativistic quantity in fully  non-linear cosmological large-scale structure studies. Traditionally, non-linear structure formation in standard $\Lambda$CDM cosmology is studied using N-body simulations, based on Newtonian gravitational dynamics on an expanding background. When one derives the Newtonian regime in a way that is a consistent approximation to the Einstein equations, a gravito-magnetic vector potential - giving rise to frame dragging - is present in the metric in addition to the usual Newtonian scalar potential. At leading order, this vector potential does not affect the matter dynamics, thus it can be computed from Newtonian N-body simulations. We explain how we compute the vector potential from simulations in $\Lambda$CDM and examine its magnitude relative to the scalar potential. We also discuss some possible observable effects. \end{abstract}

\maketitle

{\it Introduction.} Modern Cosmology is usually studied in two limits. On the largest scales, a perturbative approach is used in a general-relativistic framework. On small scales, non-linearities are treated in a Newtonian fashion, often with the use of N-body simulations.

Few attempts have been made to go beyond the Newtonian approximation on non-linear scales by including post-Newtonian type corrections \cite{tomitaflat, asada,mataterra,takfut97, carbone,hwangpn}. However, no attempt has been made to include post-Newtonian corrections in N-body simulations of cosmological large scale structure. Investigations have been carried out into the interpretation of N-body simulations on large scales, of the order of the Hubble length \cite{chis11,green12}. In \cite{green12}, they go further and examine the dictionary between Newtonian and relativistic cosmologies on all scales and how accurately Newtonian cosmology satisfy the Einstein equations. Of course, no matter how well the Newtonian dynamics capture the full GR dynamics, there are GR quantities on all scales that have no counterpart in Newtonian theory.

Recently, a new approximation scheme has been developed, dubbed the post-Friedmann approach \cite{postf}, with the aim of providing a unified framework for all scales, from the fully non-linear Newtonian regime to the largest scales where relativistic effects become important \cite{icgpaper}. It is based on an expansion in inverse powers of the speed of light, $c$, in a post-Newtonian \cite{chand} fashion, adapted to cosmology. When linearised, this approach correctly reproduces the linear general-relativistic perturbation theory. When one derives the Newtonian regime in this approach, in a way that is a consistent approximation to the Einstein equations, a vector potential must be present in the metric in addition to the usual Newtonian scalar gravitational potential.

This vector potential is non-dynamical at leading order, therefore it does not affect the matter dynamics. It is sourced only by terms that appear in Newtonian gravity, so it can be extracted from N-body simulations. Physically, this vector potential represents the gravitomagnetic field that is generally present in metric theories of gravity such as General Relativity. Its typical effect is  frame dragging, a ubiquitous relativistic effect, well known in cosmological perturbation theory \cite{bard1980} and in black hole systems \cite{thorne}. Furthermore, in the solar system, Gravity Probe B \cite{gpb} has measured the frame dragging of the Earth. The computation of the frame-dragging vector potential from Newtonian N-body cosmological simulations is the main result of this paper. It is the first time that an intrinsically relativistic quantity, i.e. a quantity with no counterpart in Newtonian cosmology, has been computed in fully non-linear cosmological large-scale structure studies.

{\it Post-Friedmann Approach} We briefly present the pertinent details of the post-Friedmann approach. For more details see \cite{postf}. The starting point of this approach is an expansion of the perturbed metric, in Poisson gauge \cite{mabert,poisgaug,wandsmal}, up to order $c^{-5}$:
\begin{eqnarray}
 g_{00}&=&-\left[1-\frac{2U_N}{c^2}+\frac{1}{c^4}\left(2U^2_N-4U_P \right)\right]\nonumber\\
g_{0i}&=&-\frac{aB^N_i}{c^3}-\frac{aB^P_i}{c^5}\\
g_{ij}&=&a^2\left[\left(1+\frac{2V_N}{c^2}+\frac{1}{c^4}\left(2V^2_N+4V_P \right) \right)\delta_{ij} +\frac{h_{ij}}{c^4}\right]\nonumber
\end{eqnarray}
Note that the background metric here is the flat FLRW metric, not the Minkowski metric, i.e.\ a standard $\Lambda$CDM cosmology is assumed. The $g_{00}$ and $g_{ij}$ scalar potentials have been split into the Newtonian ($U_N$, $V_N$) and post-Friedmann ($U_P$, $V_P$) components. Similarly, the vector potential has been split up into $B^N_i$ and $B^P_i$. Since this metric is in the Poisson gauge, the three-vectors $B^N_i$ and $B^P_i$ are divergenceless, $B^N_{i,i}=0$ and $B^P_{i,i}=0$. In addition, $h_{ij}$ is transverse and tracefree, $h^i_i=h_{ij}^{,i}=0$. Note that at this order, $h_{ij}$ is not dynamical, so it does not represent gravitational waves. From a post-Friedmann viewpoint, there are two different levels of perturbations in the theory, corresponding to terms of order $c^{-2}$ and $c^{-3}$, or of order $c^{-4}$ and $c^{-5}$ respectively. Defining ``resummed" variables, such as $\Phi=-U_N+c^{-2}\left(U^2_N-2U_P \right)$, then calculating the Einstein equations and linearising them, reproduces linear GR perturbation theory in Poisson gauge. Thus, this approach is capable of describing structure formation on the largest scales.

By retaining only the leading order terms in the $c^{-1}$ expansion, one recovers Newtonian cosmology, albeit with a couple of subtleties. The first is that the space-time metric is a well-defined approximate solution of the Einstein equations. The second is that we have an additional equation, which is a constraint equation for the frame-dragging vector  potential $B^N_i$. The full system of equations, as obtained from the Einstein and hydrodynamic equations \cite{postf,milillo:2010}, is as follows.
\begin{eqnarray}
&&\frac{d\delta}{dt}+\frac{v^i_{,i}}{a}\left( 1+\delta \right)=0\\ &&\frac{dv_i}{dt}+\frac{\dot{a}}{a}v_i=\frac{1}{a}U_{N,i}\nonumber\\
&&\frac{1}{c^2a^2}\nabla^2V_N=-\frac{4\pi G}{c^2}\bar{\rho} \delta \nonumber\\ &&\frac{2}{c^2a^2}\nabla^2\left(V_N-U_N \right)=0\nonumber\\
&&\frac{1}{c^3}\hspace{-0.1cm}\left[\frac{2\dot{a}}{a^2}U_{N,i}+\frac{2}{a}\dot{V}_{N,i}-\frac{1}{2a^2}\nabla^2B^N_i\right]\hspace{-0.1cm}=\hspace{-0.1cm}\frac{8\pi G\bar{\rho}}{c^3}\left(1+\delta \right)v_i\nonumber
\end{eqnarray}
As expected, we have the Newtonian continuity, Euler and Poissons equation from the Einstein equations, where $\bar{\rho}$ is the background matter density and $\delta=(\rho-\bar{\rho})/\bar{\rho}$ the density contrast. There is also an equation forcing the scalar potentials $V_N$ and $U_N$ to be equal, consistent with there being only one scalar potential in Newtonian theory. The final equation is the extra equation showing that, even in the Newtonian regime, the frame-dragging potential $B^N_i$ should not be set to zero in general; this would correspond to putting an extra constraint on the Newtonian dynamics.

The potential $B^N_i$ is sourced by the vector part of the energy current $\rho \vec{v}$: Taking the curl of the vector potential equation in order to remove the scalar part, we obtain
\begin{equation}
\label{eqn:curl}
 \nabla \times \nabla^2 \vec{B}^N=-\left(16\pi G \bar{\rho} a^2\right)\nabla\times\left[(1+\delta)\vec{v} \right]
\end{equation}
Note that this equation is essentially the same as the equivalent equations in \cite{takfut97,hwangnonlin,green12}. Although $B^N_i$ doesn't influence the matter dynamics at this order, it is part of the metric and will affect cosmological observables through its effect on photon geodesics. We discuss some of the possible observational consequences later. We now compute the right hand side of equation (\ref{eqn:curl}) from N-body simulations and thus construct the power spectrum of the vector potential.

We will be dealing with vector quantities, for which there are different ways to define the power spectrum. Our power spectrum for a generic vector $\vec{v}$ is defined via
\begin{equation}
\langle \tilde {\vec v}(\vec k) \ \cdot \tilde {\vec v} ^*(\vec {k^{'}}) \rangle=(2\pi)^3 \delta^3(\vec k -\vec {k^{'}})P_{\vec v}(k)
\end{equation}
Note that for a divergenceless vector, such as ${\vec B}^N$, $k^2 P_{{\vec B}^N}(k)=P_{\nabla \times {\vec B}^N}(k)$. With our Fourier transform convention, the dimensionless power spectrum for a field X is given by ${\cal P}_{X}(k)=k^3 P_{X}(k)/2\pi^2$.\\
From equation (\ref{eqn:curl}), the power spectrum of the vector potential is given by
\begin{equation}
P_{{\vec B}^N}(k)=\left(\frac{16\pi G \bar{\rho} a^2 }{k^2}\right)^2\frac{1}{k^2}P_{\delta v}(k) \rm{,}
\end{equation}
with
\begin{eqnarray}
&&P_{\delta v}=P_{\nabla \times {\vec v}}(k)+P_{\delta \nabla \times {\vec v}}(k)+P_{(\nabla \delta) \times {\vec v}}(k) \\
&&+P_{\left(\nabla \delta \times \vec{v}\right)\left(\nabla \times \vec{v}\right)}(k) +P_{\left(\nabla \delta \times \vec{v}\right)\left( \delta\nabla \times \vec{v}\right)}(k)
+P_{\left(\delta\nabla \times \vec{v}\right)\left(\nabla \times \vec{v}\right)}(k)\nonumber
\end{eqnarray}

{\it Simulations} We have run three N-body simulations with $N_{part}=1024^3$ particles and length $160 h^{-1}$Mpc, using Gadget-2 \cite{gadget2}, in order to compute the vector potential, as well as multiple additional runs with varying number of particles and box size. To allow comparison to previous studies of vorticity \cite{pueb} the simulations were run with dark matter particles only and with a cosmology $\Omega_m=0.27$, $\Omega_{\Lambda}=0.73$, $\Omega_b=0.046$, $h=0.72$, $\tau=0.088$, $\sigma_8=0.9$ and $n_s=1$. All of the simulations started at redshift 50 and had their initial conditions created using 2LPTic \cite{2lptb}.

Traditional methods of extracting fields from N-body simulations, such as cloud-in-cells (CIC) \cite{cic} have several shortcomings when applied to velocity fields. One is that the field is only sampled where there are particles, so in a low density region the velocity field is artificially set to zero. In addition, the extracted field will be a mass-weighted, rather than volume weighted field.

The Delauney tesselation of a set of points creates a unique set of tetrahedra. These tetrahedra have their nodes located at the particles' positions and the circumsphere of each tetrahedron contains no other particles. For more details on the Delauney Tesselation, see e.g. \cite{dtfe1,dtfe2}. For this work, we have used the publicly available Delauney Tesselation Field Estimator (DTFE) code \cite{dtfecode}. This code works by first constructing the Delauney tesselation and then linearly interpolating the velocities of the nodes across each tetrahedron, so the gradient of the velocity field is constant over each tetrahedron. The velocity field and its gradients are now known everywhere. In order to get a smoothed field from the simulation, a regular N$^3_{\rm{grid}}$ grid is laid down. The code then samples points at random in each grid cell and averages the results, giving a value for each grid cell. For this analysis, the code sampled 100 points per grid cell. However, varying this up to 1000 made no difference to the results.

One of the disadvantages of the tesselation code is that, unlike CIC methods, the window function cannot be easily deconvolved; the window function will be different in different regions of the simulation. We can examine the effects of the window function by varying the grid size used to analyse a given simulation. Our main result was calculated using N$^3_{\rm{grid}}=$ N$_{part}$, but varying the grid size makes no difference except on the smallest scales. The output from the DTFE code is Fourier transformed and the modulus-squared values of the transformed field are averaged in bins for given ranges of wavenumber $k$. We used N$_{\rm{grid}}$/4 bins in our analyses, however varying this value did not affect the results.
\begin{figure}[t]
\begin{center}
\includegraphics[width=2.5in,angle=270]{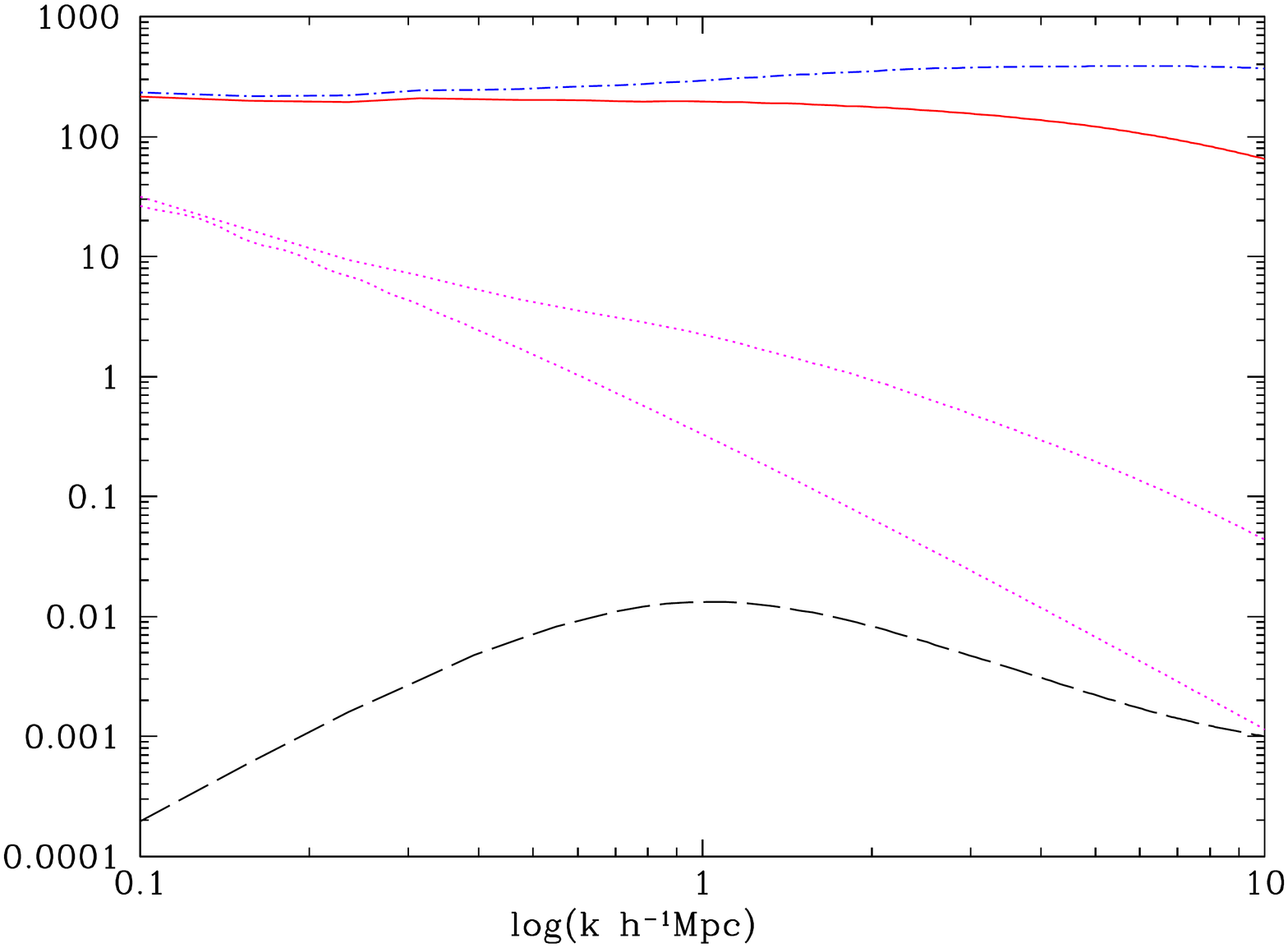}
\end{center}
\caption{The power spectra of the three source terms of the vector potential, as extracted from N-body simulations. The solid (red) line is for $\delta \nabla \times \vec{v}$, the dot-dashed (blue) line is for $\nabla \delta \times \vec{v}$, the dashed (black) line is for $\nabla \times \vec{v}$ and the dotted (magenta) lines are the linear and non-linear matter power spectra for comparison. The power spectra plotted here are given by $P(k)/\left(f^2{\cal H}^2(2\pi)^3\right)$, where ${\cal H}$ is the conformal time Hubble constant and $f=d \ln D/d \ln a$ is the logarithmic derivative of the linear growth factor $D$, see \cite{pueb}.}
\label{fig:components}
\end{figure}

{\it Convergence and Robustness}
\begin{figure}[t]
\begin{center}
\includegraphics[width=2.5in,angle=270]{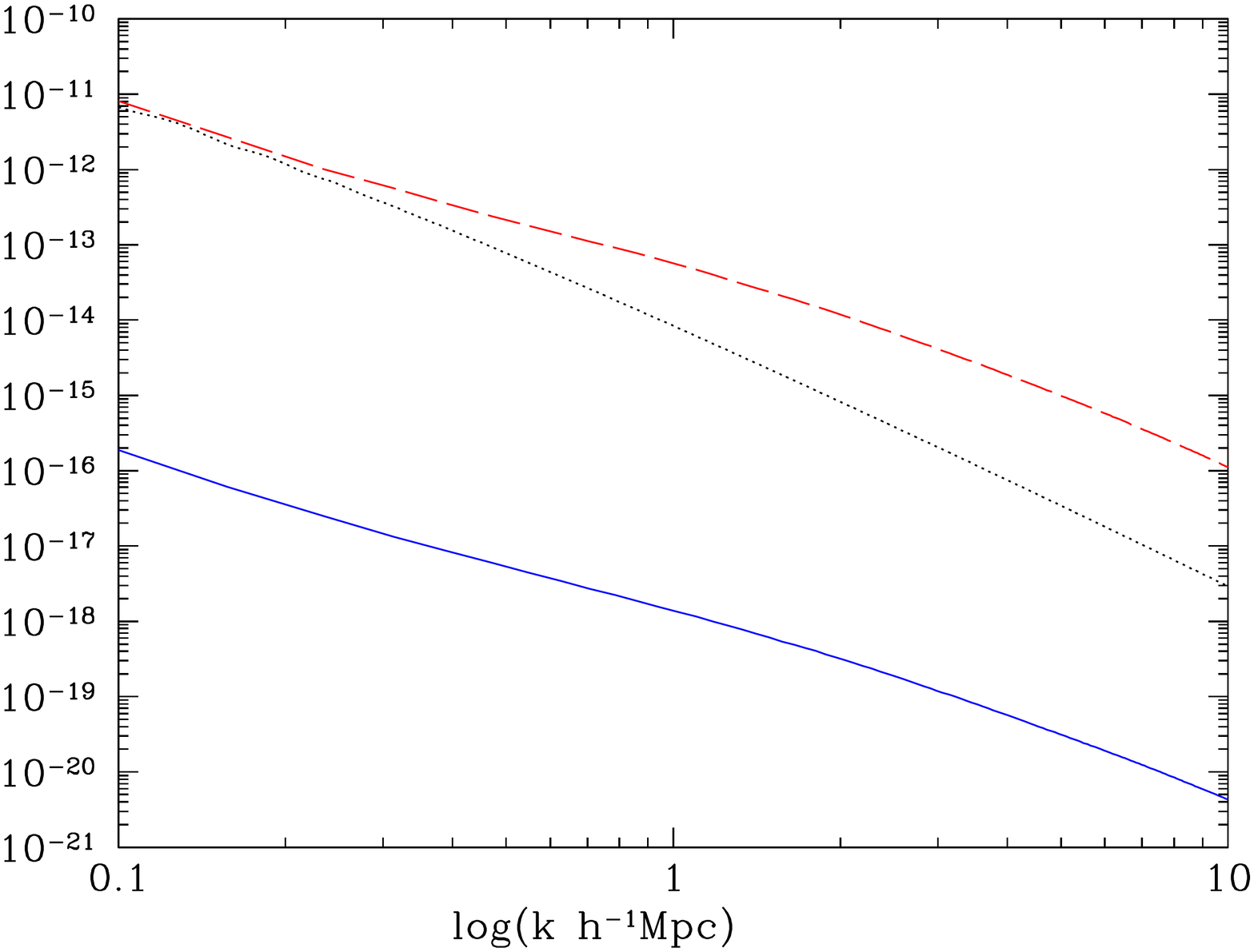}
\end{center}
\caption{The power spectra ${\cal P}_{\Phi}$ and ${\cal P}_{\vec{B}^{N}}$ of the Newtonian scalar potential (dashed red line) and the frame-dragging vector potential (solid blue line) as a function of scale, as extracted from N-body simulations. The dotted (black) line is the linear theory scalar potential for comparison. }
\label{fig:pot}
\end{figure}
Firstly, we consider some consistency checks on the extracted fields. We calculate the density power spectrum with the state-of-the-art code POWMES \cite{powmes} and check for consistency with our DTFE result. In addition, the power spectrum of the gradient of the density, which is part of one of the quantities required for the vector potential, can be extracted by itself. The power spectrum of the gradient of the density should satisfy $P_{\nabla \delta}(k)=k^2P_{\delta}(k)$, so we can check that the extraction of the two fields is consistent. A similar check can be performed for the velocity fields: As pointed out by \cite{pueb}, $k^2 P_{\vec{v}}=P_{\nabla \cdot \vec{v}}+P_{\nabla \times \vec{v}}$, so we can extract all three fields and check that they satisfy this relation. The fields do indeed satisfy this constraint, up to the smallest scales where the window function starts to have an effect. This is one way to see the effects of the window function.

We can also compare our extracted velocity spectra to \cite{pueb} where the velocity spectra were also extracted using the Delauney tesselation method. However, a different code was used that implemented the tesselation differently, see \cite{pueb} for details. For simulations with the same parameters, our extracted vorticity power spectra are consistent with this paper and show the same dependence on resolution.

A full study of the effect of box size and mass resolution on the extracted vector potential is beyond the scope of this letter. Nonetheless, for high resolution simulations that are suitable for studies of vorticity, there appear to be no significant systematic issues with resolution or box size. However, the variation amongst realisations is greater for quantities such as the vorticity, and by extension the vector potential, than for quantities such as the density and velocity divergence. This further complicates the issue and increases the required computational resources. A comprehensive study will be presented in a forthcoming publication \cite{ports2}.

{\it Results} The power spectra, averaged from three high resolution N-body simulations of length $160h^{-1}$Mpc and N$_{part}=1024^3$, of the three source terms of the vector potential, $\delta\nabla \times \vec{v}$, $\nabla \delta \times \vec{v}$ and $\nabla \times \vec{v}$, are shown in figure \ref{fig:components} alongside the linear and non-linear matter power spectra. The power spectra plotted here are given by $P(k)/\left(f^2{\cal H}^2(2\pi)^3\right)$, where ${\cal H}$ is the conformal time Hubble constant and $f=d \ln D/d \ln a$ is the logarithmic derivative of the linear growth factor $D$.  These units are chosen such that the power spectrum of the velocity divergence agrees with the density power spectrum on linear scales, following \cite{pueb}. We can see that it is the non-linear terms that are the dominant sources of the vector potential, with the vorticity contribution (the sole contribution in linear perturbation theory) being sub-dominant on all scales.

\begin{figure}[t]
\begin{center}
\includegraphics[width=2.5in,angle=270]{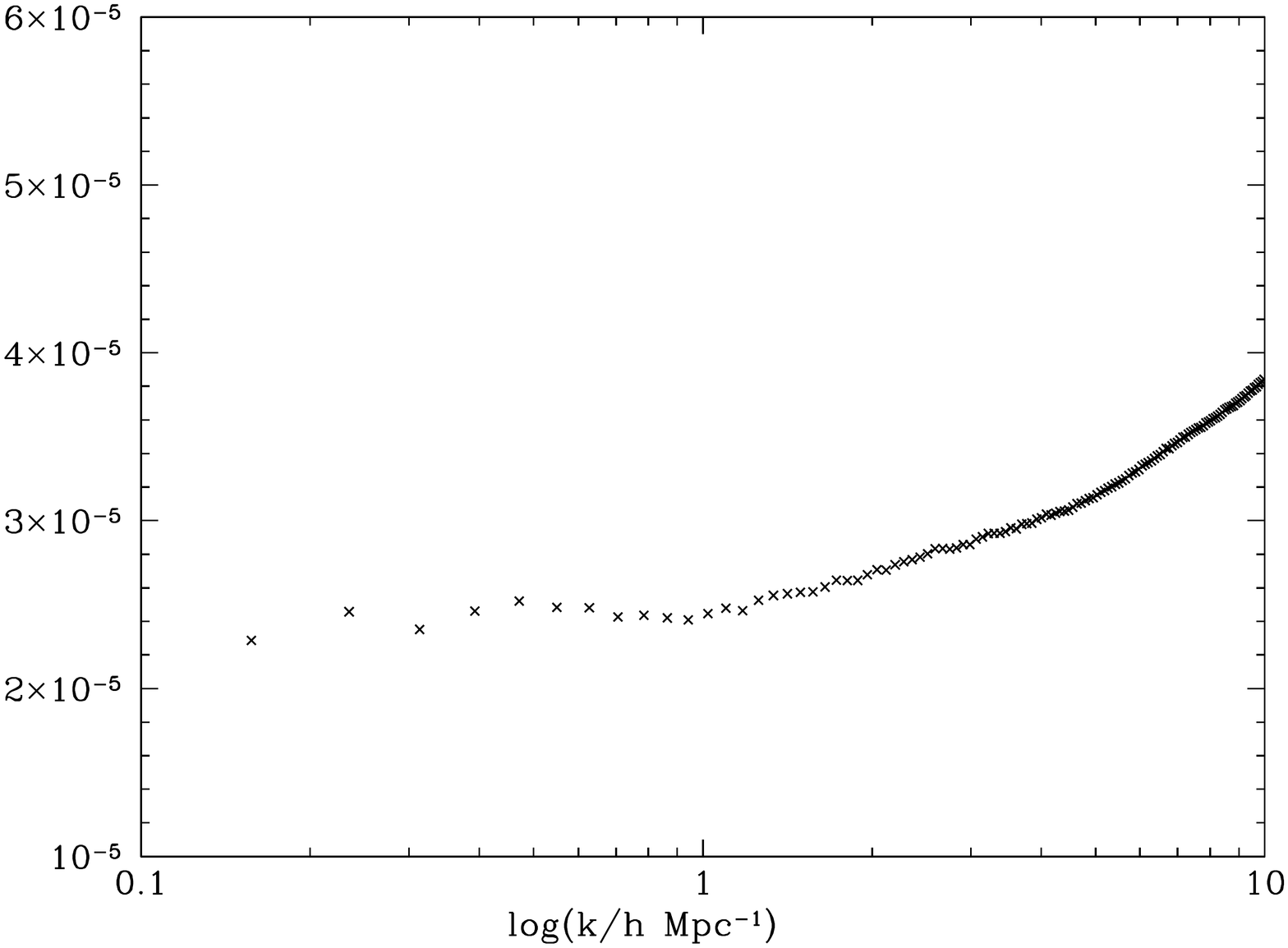}
\end{center}
\caption{The ratio of the power spectra of the Newtonian scalar potential and the gravito-magnetic vector potential as a function of scale, as extracted from N-body simulations.}
\label{fig:ratio}
\end{figure}

The average of the vector potential over the three high resolution simulations is shown in figure \ref{fig:pot}, along with the standard scalar gravitational potential. The power spectra plotted here are the dimensionless power spectra ${\cal P}_{\Phi}$ and ${\cal P}_{\vec{B}^{N}}$, as defined earlier, in natural units where $c=1$. For comparison, the linear theory scalar potential is shown as well. The ratio of the average vector potential to the average scalar potential is shown in figure \ref{fig:ratio}. These graphs show that the ratio of the vector to scalar gravitional potential is fairly constant well in to the non-linear regime, with the vector being of order $10^5$ times smaller than the scalar potential.

In \cite{0812.1349} (see also \cite{molmat}), the expected vector potential at second order in General Relativistic perturbation theory is calculated. Since this method has a different regime of validity to ours, the two methods should not be expected to agree fully. However, we can compare the qualitative behaviour in the two results. The order of magnitude of the vector potential is similar in both cases, although in our case the vector potential is larger and becomes increasingly larger on smaller scales. Over the range of overlap for the two methods, the ratio between the scalar and vector potential has a fairly constant value of order $10^{-5}$. The difference is that the vector potential in \cite{0812.1349} is of order $10^{-5}$ times the \emph{linear} scalar potential, whereas ours is relative to the full non-linear scalar potential. This similar qualitative behaviour is reassuring.

{\it Observability} A vector potential present in the metric will influence several cosmological observables. The most obvious one is weak gravitational lensing. The usual quantity considered in weak lensing is the convergence, $\kappa$, which is the isotropic expansion of a galaxy image. Here, we show how the vector potential affects the convergence power spectrum $P_{\kappa}$. We follow a treatment similar to \cite{wlstrings}, however we explicitly include powers of $c$ and work up to order $c^{-3}$ rather than using linear General Relativistic perturbation theory.

The starting point is the metric $g_{00}=-1$, $g_{0i}=-ac^{-3}B^N_i$ and $g_{ij}=a^2\delta_{ij}$. This yields a deflection angle
\begin{equation}
 \theta^{\prime}_i=\theta_i+\int^{\chi}_0d\chi^{'}\left(B^N_{\chi,i}-B^N_{i,\chi}\right)\left(1-\frac{\chi^{'}}{\chi}\right)\text{.}
\end{equation}
Compared to \cite{wlstrings}, the $\dot{B^N_i}$ term has vanished since it is order $c^{-4}$ in this expansion. The convergence is the trace of the distortion matrix $\psi_{ij}$, given by $\psi_{ij}=\partial \theta^\prime_i/\partial \theta_j-\delta_{ij}$. Following \cite{dodelson}, and working in the small angle limit with the Limber approximation, gives a convergence power spectrum
\begin{equation}
P^{B^N}_{\kappa}(l)=\frac{l^4}{8}\int^{\chi_{\infty}}_0\! \! d\chi \frac{g^2(\chi)}{\chi^6}P_{B^N}(l/\chi)
\end{equation}
where $g(\chi)$ is the weak-lensing weight function. The convergence power spectrum caused by the scalar gravitational potential is
\begin{equation}
P^{\Phi}_{\kappa}(l)=l^4\int^{\chi_{\infty}}_0\! \! d\chi \frac{g^2(\chi)}{\chi^6}P_{\Phi}(l/\chi)
\end{equation}
We can see that the vector and scalar potentials contribute in a similar fashion to the convergence power spectrum, although the vector power spectrum is much smaller than the scalar power spectrum. However, at order $c^{-4}$, the time derivative of the vector potential will generate the odd parity B-mode of cosmic shear that is not generated by the scalar potential at first order. In forthcoming work, we will examine the weak lensing power spectra up to order $c^{-4}$ and comment on the observability of the vector potential \cite{ports2}.

{\it Conclusion and discussion} The post-Friedmann approach \cite{postf} provides a framework for examining post-Newtonian effects in cosmology. The primary result of this paper is the computation of the post-Friedmann frame-dragging vector potential at leading order, i.e.\ in the Newtonian dynamical regime,  as shown in figure \ref{fig:pot}. This is the first time that an intrinsically relativistic quantity has been calculated in full non-linearity in simulations of cosmological structure formation.

For sufficiently high resolution simulations, the power spectrum of the vector potential appears to converge. The agreement of the density and vorticity fields with other methods \cite{powmes,pueb} and similar qualitative behaviour of the vector potential to analytic results \cite{0812.1349} support our numerical results. As mentioned above, although this vector potential does not influence matter dynamics at this order, it will affect photon geodesics, so the first place to look for the effects of this vector potential is in weak-lensing surveys. The large ratio between the power spectra of the vector gravitational potential and the scalar gravitational potential means that the effects of the vector potential are unlikely to be detected in the usual convergence or E-mode spectra. However, the time derivative of a vector potential generates the B-mode spectrum \cite{wlstrings}, which is not produced by the scalar potential and thus may allow the vector potential to be observed. Another effect where the vector potential may be observable is lensing of the CMB, particularly polarisation: It is known that the scalar lensing potential lenses the polarisation E-mode into the B-mode and it is possible that a vector potential is more efficient at this process, potentially rendering the vector potential observable.

The magnitude of the  vector potential we have computed also supports the validity of Newtonian N-body simulations in $\Lambda$CDM cosmology: Since the vector potential is the first relativistic addition to Newtonian theory, its small magnitude relative to the scalar potential supports the assertion that on sufficiently small scales, the relativistic corrections to Newtonian gravity are sufficiently sub-dominant. A much larger measured value would suggest that a relativistic treatment is essential for structure formation in $\Lambda$CDM. From the point of view of \cite{green12}, the small size of the vector potential suggests that the abridged dictionary, corresponding to the dictionary in \cite{chis11} can be used. Nonetheless, as shown here, even in a regime where the cosmological dynamics is Newtonian a relativistic framework is essential for the interpretation, and  relativistic effects can be computed that are potentially observable. At next order, possible effects of the non-zero difference between the two scalar potentials that appear in the  post-Friedmann approach \cite{postf,milillo:2010}, also consistently with second-order relativistic perturbation theory \cite{poisgaug,wandsmal} and other studies \cite{green12}, remain to be understood. These relativistic  non-linear effects are potentially more important in clustering/coupled Dark Energy and modified gravity cosmological models.

{\sl Acknowledgements} We thank Marius Cautun for help with the DTFE code. The simulations in this paper were run on the SCIAMA computer at Portsmouth and analysed on COSMOS at Cambridge. This work was supported by STFC grants ST/H002774/1 and ST/K00090X/1. Please contact the authors to request access to research materials discussed in this paper.
\bibliography{portsletter}
\end{document}